# Triangular nanobeam photonic cavities in single crystal diamond

**Igal Bayn[1], Boris Meyler[1], Joseph Salzman[1] and Rafi Kalish[2]**

1 Department of Electrical Engineering and Microelectronics Research Center, Technion Haifa, 32000, Israel

2 Department of Physics and Solid State Institute, Technion Haifa, 32000, Israel

E-mail: eebayn@techunix.technion.ac.il

**Abstract.** Diamond photonics provides an attractive architecture to explore room temperature cavity quantum electrodynamics and to realize scalable multi-qubit computing. Here we review the present state of diamond photonic technology. The design, fabrication and characterization of a novel triangular cross section nanobeam cavity produced in a single crystal diamond is demonstrated. The present cavity design, based on a triangular cross section allows vertical confinement and better signal collection efficiency than that of slab-based nanocavities, and eliminates the need for a pre-existing membrane. The nanobeam is fabricated by Focused-Ion-Beam (FIB) patterning. The cavity is characterized by a confocal photoluminescence. The modes display quality factors of $Q \sim 220$ and are deviated in wavelength by only $\sim 1.7nm$ from the NV$^-$ color center zero phonon line (ZPL). The measured results are found in good agreement with 3D Finite-Difference-Time-Domain (FDTD) calculations. A more advanced cavity design with $Q=22,000$ is modeled, showing the potential for high-Q implementations using the triangular cavity design. The prospects of this concept and its application to spin non-demolition measurement and quantum computing are discussed.

## 1. Introduction

Quantum Information Technology (QIT) may revolutionize computing and communication by allowing solution of problems that nowadays computers are inefficient in, such as non-polynomial (NP) algorithms, ultimately secure communication, and modeling of many-body systems [1]. The negative nitrogen-vacancy color center ($NV^-$) in a single crystal diamond is an attractive candidate for solid state QIT [2],[3]. The $NV^-$ ground spin-triplet state can be read-out optically, manipulated by microwave and polarized at room temperature. In addition, it shows extremely long life and coherence time ($T_1$ and $T_2$). These make $NV^-$ a promising platform for quantum information storage and computation [2]. Single and two-qubit operations with $NV^-$ were demonstrated by several groups [4]-[7], however, they relied on a dipole-dipole mechanism operating at distance lower than *50nm*. Since controlled creation of $NV^-$ to form interacting qubits in a certain location is beyond today's technological capabilities, these successes in quantum computation shall by attributed to an extensive search procedure in a "hunt" for adequate color center/nuclei configuration. Since the interacting elements are randomly distributed in space, extending this approach to many-qubit manipulation is expected to be a challenge [8]-[10]. Therefore, instead of search for an adequate qubit configuration, an architecture which will allow *controlled* interaction between certain qubits will be beneficial. In addition, quantum processing imposes two contradictory requirements: the qubits must interact to perform the computation, while their pre- and post-computation interaction shall be reduced to minimum to allow efficient quantum data storage. To accomplish this quantum architecture, the following conditions shall be satisfied : (1) qubits shall be positioned sufficiently far from each other to reduce inevitable decoherence (the actual length-scale will be set by the photonic design discussed below); (2) the interaction shall be carried by photons, which is the practical way to implement many-qubit entanglement essential for quantum algorithm realization; (3) there must be a mechanism to switch-off the qubit-qubit interaction, allowing efficient data storage in the pre- post-computation periods. (See for example [11]). These conditions if satisfied, allow an ultimate increase in the number of $NV^-$ involved, thus paving a way to a true large-scale-integration of QIT. One attractive way to implement this is in a photonic architecture, where each qubit (the $NV^-$ site in diamond) is registered within a high-Q cavity, and the cavities are optically interconnected via waveguides [2], [12],[13]. The very convenient choice for implementing such architecture is in a photonic crystal configuration, where the cavity and waveguide structure can be defined and fabricated, thus allowing the interconnection between certain qubits . The control over NV$^-$ coupling to the cavity and as a result over the interaction with another qubit via the waveguide may be realized by Stark effect [14]. The single-qubit transition between the upper and the lower spin state is possible by applying microwave radiation pulses at *2.88GHz*. These produce Rabi oscillations between $m_s=0$ and, for instance, $m_s=-1$ states (Note that the quantum state is encoded either into $m_s=1$ or $m_s=-1$, while the degeneracy between them may be relieved by applying a static magnetic field. Here, we have chosen the *-1* state).

In the field of diamond photonics major progress has been made in recent years. Computational efforts led to ultra-high-*Q* nanocavity designs [15]-[19]  a nanocrystal high-*Q* cavity was demonstrated [20]. In addition, single nanocrystal diamond NV$^-$ coupled to a gallium phosphide photonic crystal cavity has been realized [21]-[23]. However, it should be mentioned that single crystal diamond, rather than nanodiamond would provide significant advantage in reducing optical losses and scattering effects [20]. A photon source carved in a single-crystal diamond in the form of vertical nanowires has been fabricated and characterized, thus showing the capability for precise patterning and deep ion-etching with standard lithography tools [24]. The nanowire geometry provides a non-cavity based platform for coupling NV$^-$ emitted photons to antenna without vertical mode confinement, thus presenting a solution that does not suffer from fabrication problems discussed below. However, mode leakage into the substrate limits potential *Q* of this system. The natural conclusion to be drawn from these early experiments is that the realization of interconnected high-*Q* cavities for QIT should be implemented on single crystal diamond, in the form of a 2D-slab photonic crystal, thus requiring the formation of a free-standing membrane.

Diamond membranes were formed by a technique based on ion implantation damage. The diamond surface is irradiated with He$^+$ ions to produce a damage profile with precise dose *vs.* depth controlled distribution. The implanted sample is annealed at elevated temperature.  In regions where the damage is lower than certain threshold, the diamond structure is recovered, while in the high damage layer, the diamond is graphitized, forming a sacrificial layer, which can be chemically removed. Since the sacrificial layer is buried under the low damage region, its etching leaves a free standing diamond membrane [25],[26]. The results of such process are shown in figure 1(a).

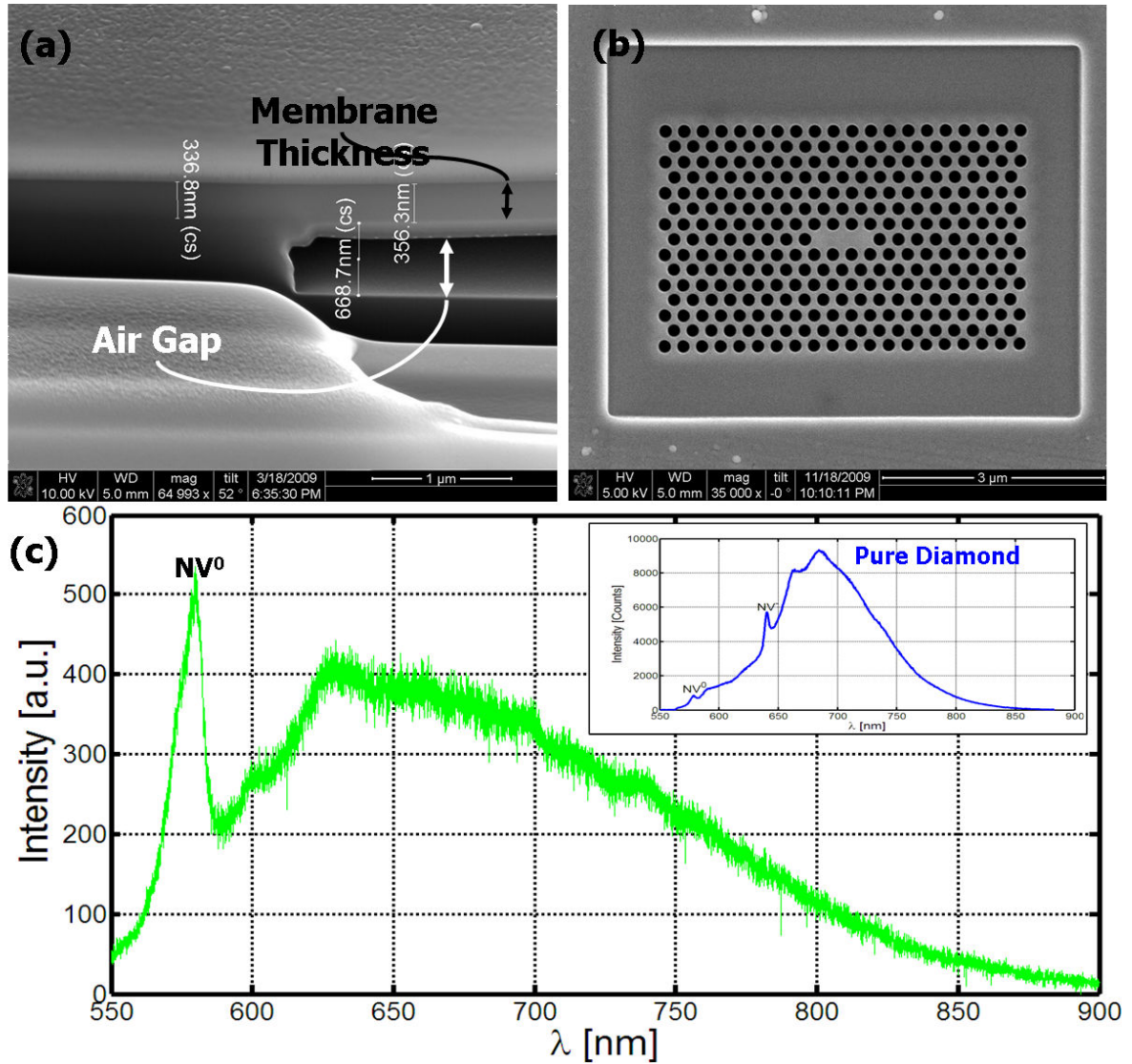

**Figure 1.** (a) Diamond membrane realized by ion implantation. (b) Diamond PC nanocavity formed by FIB [26]. (c) Spectrum modification resulted from the implantation process. In the inset unprocessed diamond spectrum [27].

The membrane fabrication technology showed that for a thick membrane (with a thickness higher than ~$1\mu$), the optical quality can be high [27]. However, for ~$200nm$ thick membrane needed for photonic crystal realization (see figure 1(b)) the unpatterned membrane displayed extreme optical losses and significant modification of the color center luminescence, as one can observe from figure 1(c) [26]-[28]. The disappearance of the NV$^-$ can be explained by residual damage due to the ion implantation causing quenching of the color center [29].The optical degradation in the case of ultra-thin membranes is believed to be related to the ion-implantation technique utilized in the membrane formation process. Until this technological obstacle is overcome, an alternative cavity design in a single crystal diamond, with potential for a high $Q$ is required. One possible solution is presented next.

**2. Triangular Nanobeam Double Barrier Cavity**

We consider first a one-dimensional structure (1D), in the form of a waveguide along the *z*-axis. The cross section of the waveguide is of sub-micron size, and physically separated from the substrate (thus forming a "bridge" or a "beam"). We denominate the structure a "nanobeam". A 1D photonic crystal is formed along the nanobeam by a periodic array of grooves (a 1D Bragg reflector, with periodicity *a*). A cavity is formed in the nanobeam by a local variation of the lattice constant, *a*, along the beam. In this way, along the *z* direction the cavity mode is confined by two *1D* photonic crystal reflectors while in *xy* plane the mode confinement relies on total internal reflection (TIR). Nanobeam cavities were developed for various material systems, such as silicon [30],[31], silicon nitride ($Si_3N_4$) [32],[33], silicon dioxide ($SiO_2$) [34]. The nanobeam was produced in those by patterning a pre-existing membrane and therefore has a rectangular cross-section, while the 1D PC defined by circular or rectangular holes with a constant spacing. Since the formation of a membrane on single crystal diamond is to be avoided due to its immature process technology, we seek an alternative geometry allowing the formation of a nanobeam separated from the substrate. One way to solve this problem is via nanobeam membrane formation either by side milling or relocation producing a rectangular cross-section as is discussed in [19]. Another possible option, the triangular-shaped cross-section nanobeam cavity design approach is introduced herein.

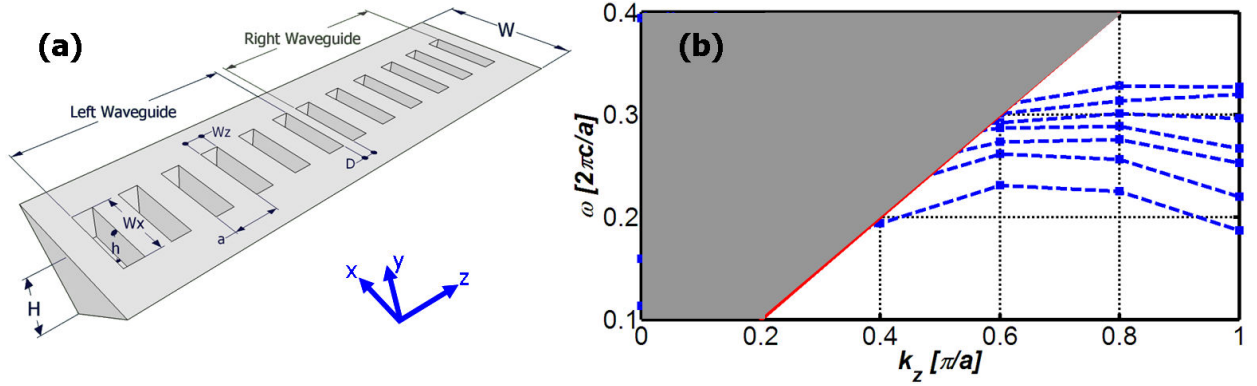

**Figure 2.** (a) Schematics of the structure of the Triangular-shaped cross section nanobeam cavity design. (b) The dispersion curves of the 1D nanobeam with following geometry: *H=2.36a, W=6a, $W_x$=0.4a, $W_y$=0.37W*.

The proposed geometry is shown schematically in figure 2(a). The beam has a Triangular-shaped cross section in the *xy* plane, and is directed along *z*. This profile supports mode guiding and induces preferential radiation into the upward direction. The beam is patterned with a series of rectangular grooves. The distance between the grooves is defined as the lattice constant *a*. The beam width and height are *W* and *H*, while the hole width and length are $W_x$ and $W_z$. The hole depth is defined as *h* and it amounts to half of beam thickness (*H/2*). In Fig. 2(b), the waveguide dispersion curves calculated by a 3D Finite Difference Time Domain (3D-FDTD) for the triangular cross section beam is depicted. The grooves carved into the nanobeam induce an upward frequency shift of the modes, due to the reduced effective refractive index, producing a "gentle" waveguide based confinement.

*2.1 Nanobeam Fabrication*

All fabrication processes, excluding essential debris etching, were performed *in situ* by Focused ion Beam (FIB) patterning (30 keV Ga ions), in a Strata 400 STEM Dual Beam (SEM/FIB) machine (FEI). Type-Ib diamond of Sumitomo was used. The abundance of nitrogen and consequently high concentration of NV centers produce a broad photoluminescence (PL) spectrum *(570÷800nm)*. This allows the confocal micro-photoluminescence spectroscopy-characterization discussed below. FIB milling strategies have been developed for the last decades in the view of the transmission electron microscopy (TEM) sample preparation [35]-[37]. In diamond these strategies and lift-off techniques were applied in the formation of several photonic devices [19], [27]. FIB precise angular stage control enables milling at an arbitrary angle ($\theta$) to the sample normal ($0°<\theta<52°$), as is shown in figure 3(a).

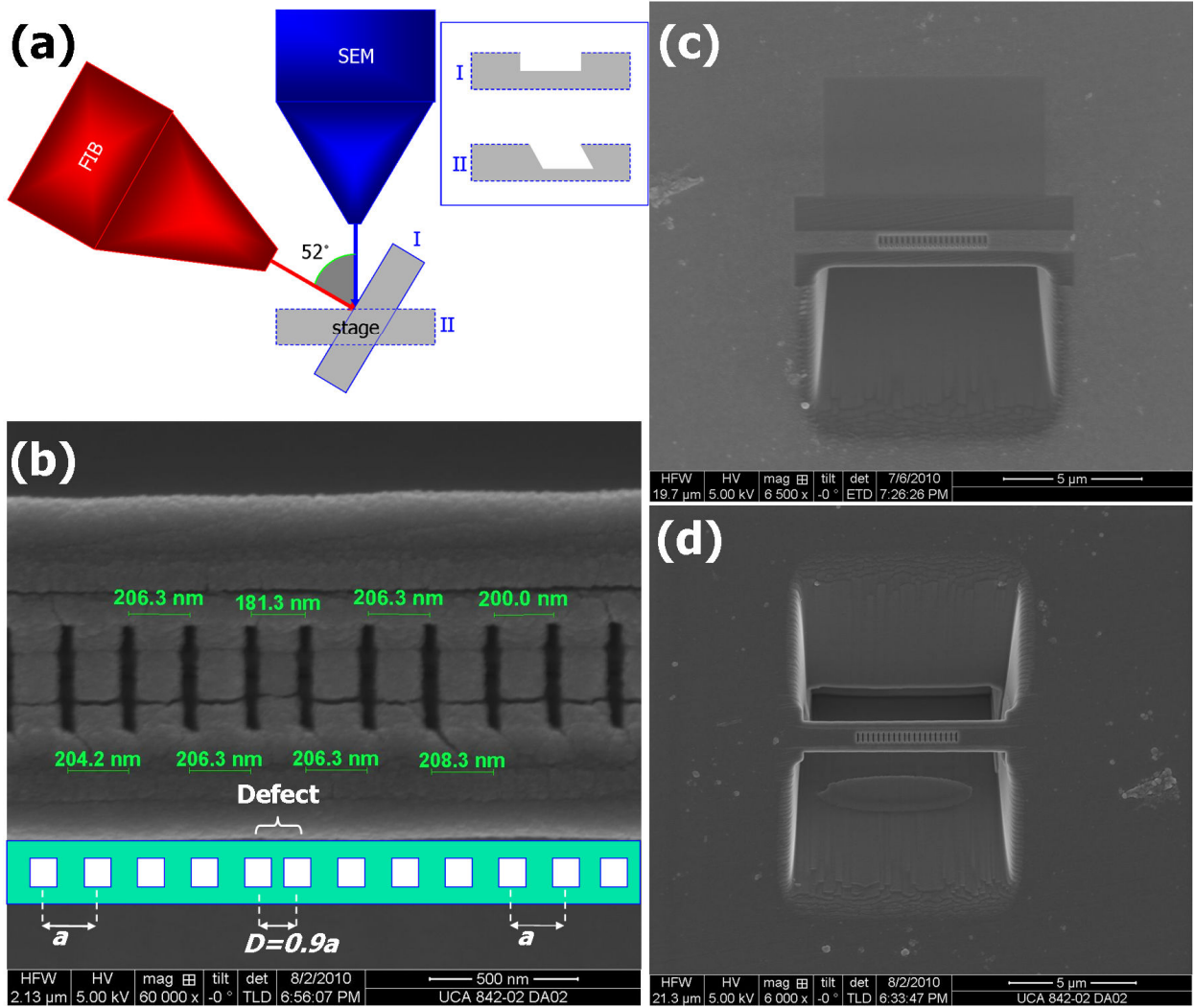

**Figure 3.** (a) Schematic of the dual-beam machine (FIB) for simultaneous patterning and electron microscopy. The carving of diamond is performed in one of two angular stage positions (I,II) at the angles of 0˚ and 52˚, respectively. (b) The defect region in the cavity. (c) Coarse carving of the triangular cut profiles by high ion beam currents. (d) Final free standing beam structure after triangular surface polishing with low current.

Here, the nanobeam grooves with a vertical cut at $\theta=0$ are milled at low current (*98pA*) (see figure 3(b)) and the triangular cut at $\theta=\pm52°$ is formed initially at a high FIB current (*2.8nA*) (see figure 3(c)). The thickness of the removed material around the nanobeam defines the vertical separation from the substrate, which amounts to $H_s\sim5\mu$. Then, the edges of the triangular-shaped profile nanobeam are polished by a low beam current (*98pA*). In this process the final beam dimensions are defined (figure 3(d)). Eventually, the debris produced by the milling process redeposit close to the milling location, are removed by etching in a boiling $HClO_4/HNO_3/H_2SO_4$ (1:1:1). The detailed nanobeam parameters are summarized in table 1. Note that after chemical etching the FIB implanted Ga ion resides near the milling surface. Secondary Ion Mass Spectrometry (SIMS) shows that the maximum concentration of $\sim 10^{22}$ *ions/cm³* appears in the depth of *1-5nm*. Then, the concentration rapidly decays with the depth increase (at the depth of *30nm* the decrease is by *3* orders of magnitude). The influence of this implanted layer on photoluminescence is discussed bellow. The ways of reducing this Ga concentration are under study.

**Table 1.** Nanobeam geometry parameters

| Lattice Constant | Defect | Beam Parameters | | | Waveguide Grating | | | Well | |
|---|---|---|---|---|---|---|---|---|---|
| $a$ | $D$ | $H$ | $W$ | $L$ | $W_x$ | $W_y$ | $h$ | $H_s$ | $W_s$ |
| 1 | 0.9a | 2.364a | 6a | 39a | 0.4a | 0.37W | 0.5H | 24.4a | 29.3a |
| 205nm | 180nm | 450nm | 1.24μ | 8μ | 82nm | 460nm | 225nm | 5μ | 6μ |

*2.2 Nanobeam Cavity* Characterization

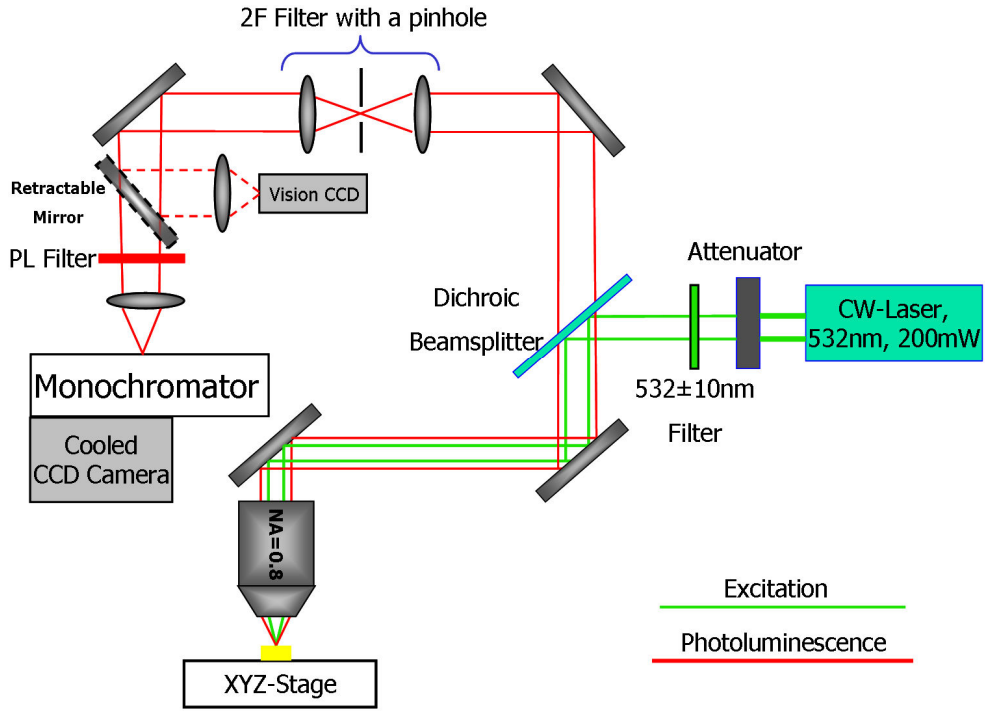

**Figure 4**. Confocal Photoluminescence set-up. The set-up consists of CW green laser, high-NA objective, piezo-motorized stage, spatial confocal filter, the excitation clean-up and long pass PL filter. These are augmented by the monochromator spectrum and vision camera.

The nanobeam cavity is characterized via a confocal photoluminescence set-up as shown in figure 4. A CW laser with a central wavelength of *λ=532nm* and output power of *200mW* is attenuated to *20mW* and filtered by a narrow band (*Δλ=10nm*) green filter. The beam is reflected from a dichroic mirror and excites the sample through an Olympus objective with *NA=0.8* and *WD=3.4mm*. The beam spot position on the sample is controlled by a 3D piezo-motorized stage. This allows precise excitation of a cavity region and performance of spatial scan. The photoluminescence (PL) signal ranges from 570 to 800nm (typical to the Type-Ib diamond, see figure 5(a)) and modified by the cavity peaks. A confocal double-*f* pinhole system produces an improved spatial resolution. At this stage, the signal either enters the PL long-pass-filter (which totally removes the excitation), fed into the monochromator for spectra read out by a cooled CCD camera or diverted via a retractable mirror into a vision CCD to observe the excitation point.

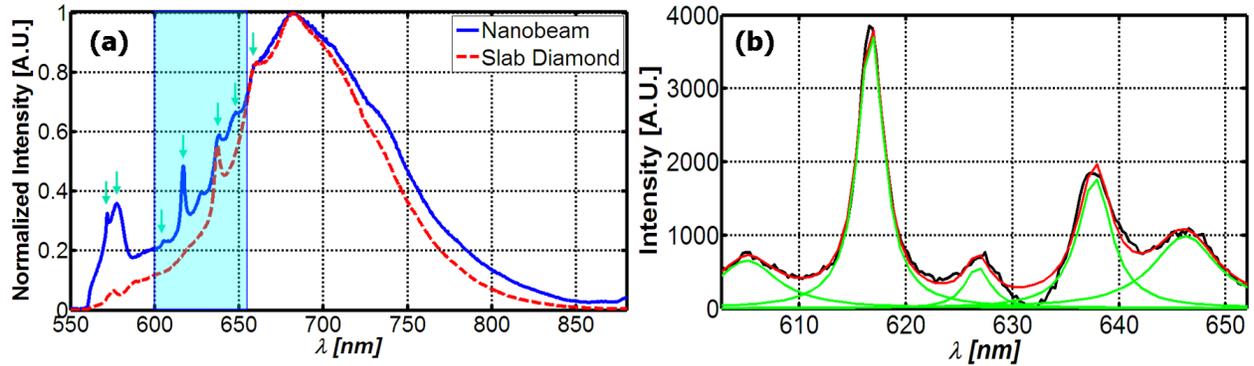

**Figure 5:** The measured cavity spectrum. (a) The cavity spectrum is marked in blue, while the unprocessed diamond is in red. The peaks are marked with cyan arrows. (b) Lorentzian fit for the peaks in the region of interest.

The confocal PL spectrum taken from the nanobeam center is shown in figure 5(a). In the same figure the spectrum of the unprocessed diamond is depicted as a reference. We note that large vertical separation ($H_s$) assures that the spectrum is obtained solely from the beam without being affected by the substrate radiation. The peaks on cavity spectrum are marked with cyan arrows. The two lower peaks at the wavelengths of *571.8* and *577.1nm* may originate from the Ion-Beam implantation, which in accordance to our previous work may modify the spectrum [27]. Moreover, for thin nanobeams the *577.1nm* peak significantly increases, thus supporting the Ion-Beam peak origin. The nature of these peaks requires further investigation. The highest peak (*658.8nm*) coincides with the knee in the unprocessed diamond spectrum, thus being disregarded in further analysis. Therefore, only the peaks at the region from *602* to *652nm* are analyzed. The Lorentzian fit of these peaks is presented in figure 5(b). From this analysis the central wavelengths and quality factors are derived.

**Table 2.** Nanobeam mode analysis. The measured modes are compared to the FDTD calculations and grouped in accordance to the $H_y$ field component symmetry (even-E or odd-O) in regard to *x* and *z* axes. The $H_y$ field component and the electric field energy $U_E$ are calculated for each mode in the *y=0* and *z=0* planes, respectively. In the bottom lines the calculated and measured values of mode wavelength and quality factors are presented.

| *Modes* | *OE1* | *OE2* | *EO1* | *EO2* | *EE1* |
|---|---|---|---|---|---|
| *Symmetry* | *OE* | *OE* | *EO* | *EO* | *EE* |
| $H_y$ | 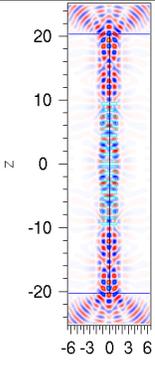 | 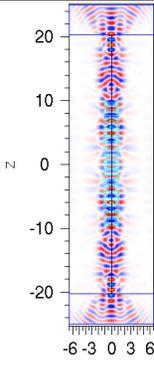 | 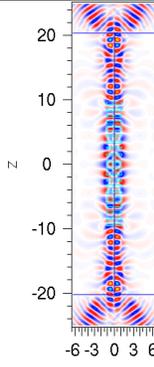 | 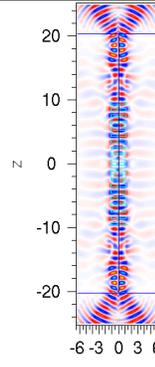 | 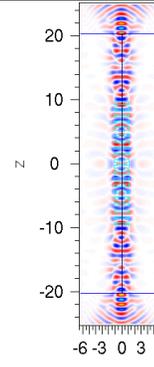 |
| $U_E\|_{z=0} = \frac{1}{2}\varepsilon\|E\|^2$ | 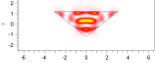 | 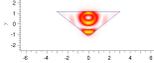 | 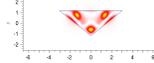 | 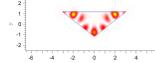 | 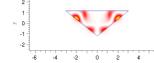 |
| $\lambda_{calculated}$ | 585 | 610 | 617.7 | 632.2 | 648 |
| $\lambda_{measured}$ | 605.4 | 616.9 | 627.4 | 638.6 | 649.6 |
| $Q_{calculated}$ | 80 | 180 | 274 | 71 | 185 |
| $Q_{measured}$ | 87 | 213 | 221 | 170 | 87 |

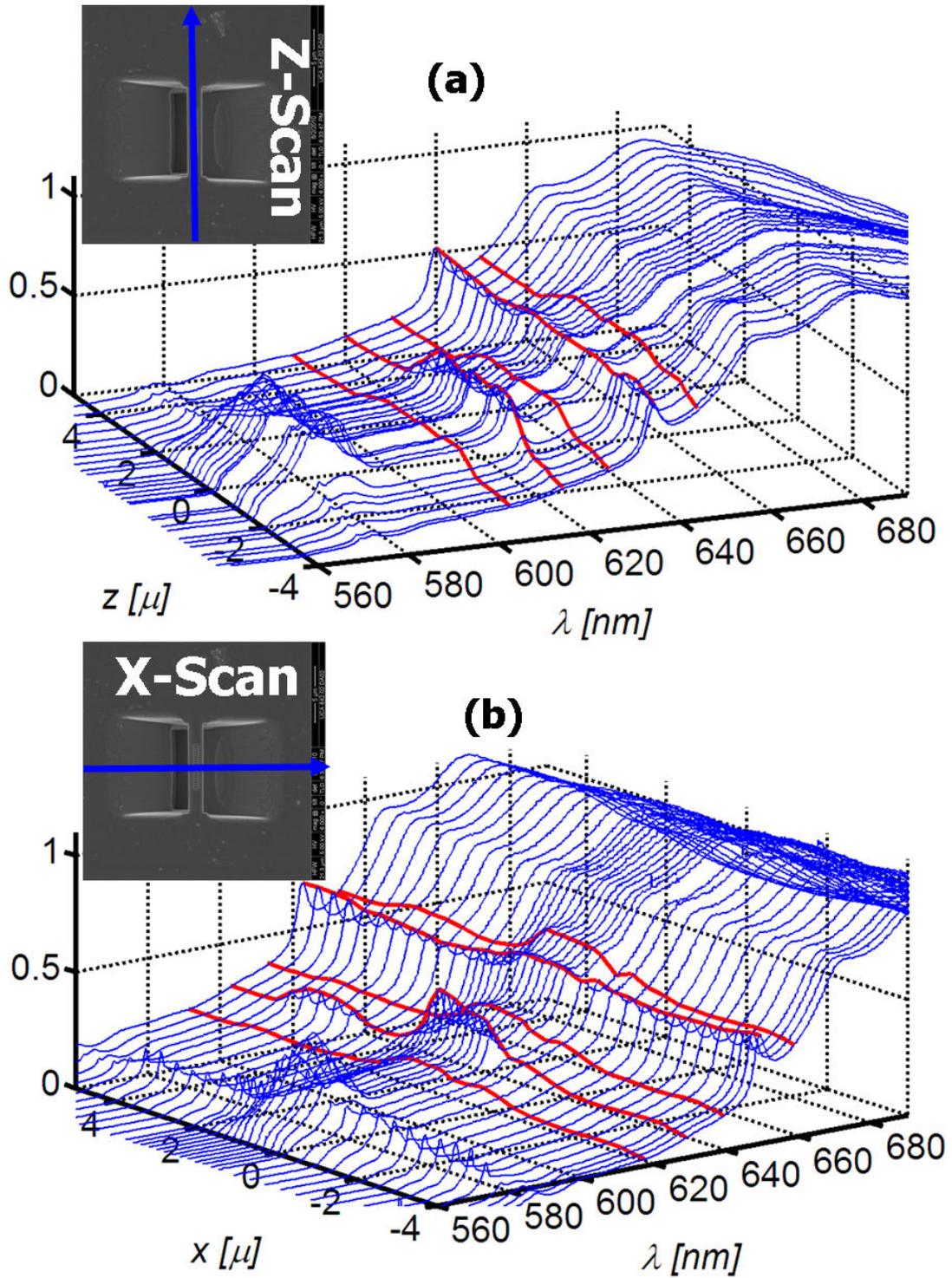

**Figure 6.** Spatial Confocal scan. The peaks in the region of interest are marked with red lines. The scan direction is shown in the figure insets. (a) Spectra scan along *z* axis across the cavity center (b) Spectra scan along *x* axis across the cavity center.

In table 2 the measured modes are summarized and compared to the Finite-Difference-Time-Domain (FDTD) calculations. The modes are divided by the magnetic field component ($H_y$) symmetry (even(E) or odd(O)) along *x* and *z* axes. The calculated profiles appear in Table 2 [38]. As one can observe, the mode wavelengths are predicted with precision better than *20nm* and this precision improves with the increase in wavelength. The quality factors are within the same order of magnitude as predicted by FDTD calculations, while the highest *Q* of *221* is obtained for the EO1 mode at the wavelength of *617.7nm*. This value is very close to the predicted one (*Q=274*). The wavelength of the mode EO2 is *638.6nm* which is only *1.7nm* higher then the NV⁻ ZPL wavelength (*637nm*). This peak presents the *Q=170* which is adequate for the realization of a single-spin room temperature read out [39].

In figure 6 the spatial scan of the spectrum along two axes of the nanobeam is presented. As one can observe, all the peaks appear at the cavity region and disappear towards the cavity edges where, naturally, unperturbed diamond spectra are obtained. Similar confocal scan has been performed in the vertical direction (*y*) to validate that the PL signal originates from the beam and is not resulted from the parasitic substrate related signal.

## *2.2 High Q nanobeam cavity*

Experimental results of the triangular nanobeam cavity in a single diamond thus exhibit low values of *Q*. In order to achieve the strong coupling regime the cavity has to exhibit values of *Q* higher than $3 \times 10^3$ [40]. To achieve this, we have modeled a modified version of the defect region in the nanobeam composed of 11 decreased lattice constant periods (*D=0.9×a*) and the thickness *H=1.65a*, as is shown in figure 7(a) The *Odd-Even-symmetry* mode (OE) profile is shown in figure 7(b, c). This cavity mode FDTD calculation shows a *Q=22,400* and $V_m=1.8 \times (\lambda/n)^3$ resulting in a Purcell factor of $F_{cav}=950$. This cavity design potentially enables to realize strong coupling regime essential for the advanced quantum computation scheme discussed in [2],[12],[13]. The ultra-high-*Q* nanobeam design with $Q \sim 10^6$ is further developed in [41]. Note that nanobeam architecture for cavity-cavity coupling needed for multi-qubit quantum computation scheme is less obvious than in the 2D-slab photonic crystal architecture, and is being considered for future research.

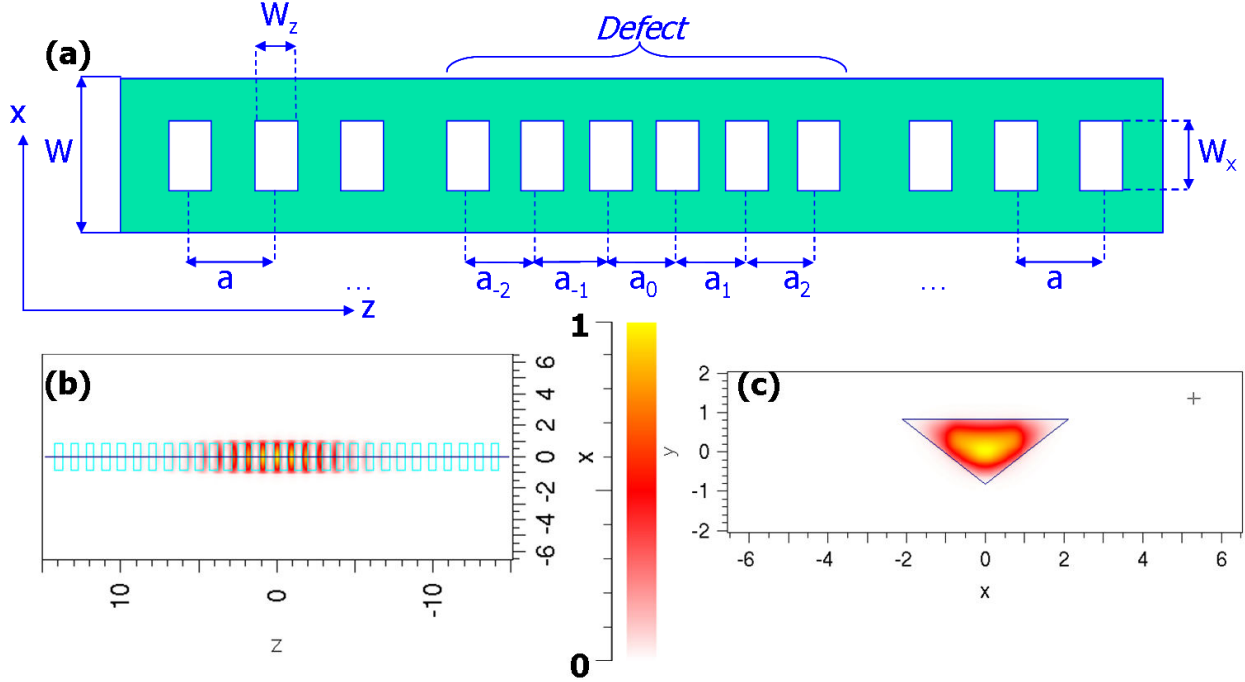

**Figure 7.** (a) Schematic design of the high-$Q$ nanobeam cavity $a_i=D=0.9A$ for $|i|<5$. (b,c) Total energy $(U = U_E + U_M = \frac{1}{2}\varepsilon|E|^2 + \frac{1}{2}\mu|H|^2)$ density profiles in $y=0$ and $z=0$ planes.

## 3. Nanobeam Applications to quantum computation schemes

The realization of a low $Q$ cavity coupled to an NV$^-$, described here, will allow fluorescence measurements to detect a single-spin. Ideally, narrow band laser excitation of $m_s=0$ ground state will produce a train of pulses characterized by a spontaneous emission time $\tau\sim 13ns$, while for $m_s=\pm 1$ a similar excitation is shelved at the $^1A$ state producing a dim period characterized by a non-radiative decay of $t=250ns$. This view relies on 100% of the emitted photons being collected. If practical collection efficiency of about *1%* is considered the dim periods become indistinguishable. When the NV$^-$ is registered into a low-$Q$ cavity ($Q\sim 200$), the detection process benefits simultaneously by two factors: First, all photons are emitted into the cavity mode with a preferential emission into the upward direction. As a result of that, the collection is improved by an order of magnitude since in triangular cross-section nanocavity an upward power flux in the $+y$ direction may be *20* times higher than the lower power flux in the $-y$ direction [41]. Second, the spontaneous emission rate of the radiative transition in the cavity $\Gamma$ is modified by the Purcell effect [42]:

$$\Gamma = F_{cav}\Gamma_0 = \frac{3}{4\pi^2}\frac{\lambda_0^3}{n^3}\frac{Q}{V_m}\Gamma_0 \qquad (1)$$

where $n$ is slab refractive index, and $\Gamma_0$ is the bulk spontaneous emission rate. This weak coupling causes a significant increase in the spontaneous emission rate (considering $V_m=2\times(\lambda_0/n)^3$ and $Q\sim200$, $F_{cav}\sim7.6$), while leaving the shelving time intact. Therefore, spin state measurement can be performed in relatively small number of excitation cycles, while its visibility is improved by a factor of ~76 (considering the collection efficiency). The detailed analysis of this non-demolition measurement for the waveguide cavity integrated scheme is given at [38][39]. In addition to the single-spin characterization, the weak coupling regime can tremendously benefit single-photon sources for the same reasons. Thus, realization of relatively low $Q$ cavity coupled to a single NV$^-$ center has an immediate impact on quantum non demolition measurement and quantum devices.

## 4 Summary

We have introduced a novel nanocavity design based on the triangular cross section nanobeam. Low $Q$ cavities were fabricated on single crystal diamond. The fabrication process avoids the detrimental effect of ion implantation damage to form a membrane. This concept can, in principle, be realized via electron-beam lithography followed by dry etching, enabling a better precision in the patterning. . The nanocavity shown here was fabricated on Type-Ib diamond enriched with nitrogen using FIB. Further control of the cavity fabrication process and a deeper study of the fabrication method (FIB) influence on a single NV center spin coherence is needed before a single color center coupled to the nanobeam resonator is produced.

**Acknowledgements**

Partial support of the Russell Berrie Nanotechnology Institute at the Technion and German Israeli Foundation (GIF) is acknowledged.